\documentstyle[12pt,aps]{revtex}

\begin{document}

\hsize = 6.5in
\widetext
\draft
\tighten
\topmargin-48pt
\evensidemargin10mm
\oddsidemargin10mm

\preprint{EFUAZ FT-96-28}

\title{On the existence of additional solutions
for equations in~the~$(1/2,0)\oplus (0,1/2)$ representation
space\thanks{Submitted  to ``Int. J. Theor. Phys."}}

\author{{\bf Valeri V. Dvoeglazov}}

\address{
Escuela de F\'{\i}sica, Universidad Aut\'onoma de Zacatecas \\
Antonio Doval\'{\i} Jaime\, s/n, Zacatecas 98068, ZAC., M\'exico\\
Internet address:  VALERI@CANTERA.REDUAZ.MX}

\date{July 15, 1996}

\maketitle

\bigskip

\begin{abstract}
We analyze dispersion relations of the equations recently proposed
by Ahluwalia for describing neutrino. Equations for type-II spinors are
deduced on the basis of the Wigner rules for left- and right- 2-spinors
and the Ryder-Burgard relation. It is shown that equations contain
acausal solutions which are similar to those of the Dirac-like
second-order equation. The latter is obtained in a similar way, provided
that we do not apply to any constraints in the process of its deriving.
\end{abstract}

\pacs{PACS numbers: 03.65.Pm, 12.90.+b}

\newpage

Recently, Ahluwalia proposed the new wave equation for describing
self/anti-self charge conjugate states $\lambda^{S,A} (p^\mu)$
of any spin~\cite{DVA96}:
\begin{eqnarray}
&&{\cal D} \lambda (p^\mu) =\nonumber\\
&&\pmatrix{-\,\openone & \zeta_\lambda\,\exp\left(
{\bf J}\,\cdot \bbox{\varphi}\right )
\,\Theta_{[j]}\,\mit{\Xi}_{[j]}\, \exp\left( {\bf J}\,\cdot
\bbox{\varphi} \right )\cr \zeta_\lambda\,\exp\left(-\, {\bf
J}\,\cdot\bbox{\varphi}\right)
\,\mit{\Xi}^{-1}_{[j]}\,\Theta_{[j]}\, \exp\left(- \,{\bf
J}\,\cdot\bbox{\varphi} \right) & -\,\openone}\,\lambda
(p^\mu)\,=\,0\,.\label{genweq1}
\end{eqnarray}
Analogous equations for $\rho^{S,A} (p^\mu)$ bispinors have been
derived in ref.~[2d]. In the $j=1/2$ case  spin matrices ${\bf
J}$ are chosen to be the Pauli matrices $\bbox{\sigma}/2$; in the
$j=1$ case, the Barut-Muzinich-Williams matrices;
$\bbox{\varphi}$ are the parameters of the Lorentz boost.  The notation
coincides with that of refs.~\cite{DVA96,DVO95a}. While formally the
$j=1/2$ equation ``may be put in the form $(\Gamma^{\mu\nu} p_\mu p_\nu
+m\Gamma^\mu p_\mu -2 m^2 \openone) \lambda (p^\mu)=0$ ... it turns out
that $\Gamma^{\mu\nu}$ and $\Gamma^\mu$ do not transform as Poincar\`e
tensors." Other forms of neutrino equations have been presented in
refs.~\cite{Ziino,DVO95a,DVO95b} and gauge interactions have been
introduced there.\footnote{The question of equivalence of these equations
still deserves further elaboration and this paper presents a certain part
of this analysis.} These constructs give alternative
insights in neutrino dynamics, which could be different from that based
on the common-used Weyl massless equation.  Indications that neutrino may
not be a Dirac particle and may have different dynamical features
have appeared
in analyses of the present experimental situation~\cite{NO}.  Earlier
considerations of this problem can be found in
refs.~\cite{Maj,Markov,MLC,Fush}.

Both the equations (\ref{genweq1}) and equations of
ref.~\cite{DVO95a,DVO95b,DVO95c} have been obtained by using different
forms of the Ryder-Burgard
relation~\cite{RB,BWW,DVA96,DVO95a,DVO95b,DVO95c} that connects
zero-momentum $(0,j)$ left- and $(j,0)$ right- spinors, and the Wigner
rules for their transformations to the frame with the momentum ${\bf p}$.
The Dirac equation may also be obtained in such
a way~\cite[footnote \# 1]{DVA96}.  The detailed
discussion of this techniques can be found in~\cite{DVO95d}.
It was claimed in ref.~\cite{DVA96} that $\lambda^S (p^\mu)$ spinors
answer for ``{\it positive} energy solutions, \ldots [meanwhile],
$\lambda^A (p^\mu)$ are the {\it negative} energy solutions".  We, in
fact, used this interpretation in~\cite{DVO95a}. Let us now check by
straightforward calculations, what dispersion relations has the equation
(\ref{genweq1}) in the case of $j=1/2$? Rewriting it to the form (31) of
ref.~\cite{DVA96} yields  the equation of the second order in $p_0$ and
the matrix in the left side has the dimension four. So, one should have
eight solutions. The analytical calculation system MATEMATICA 2.2 yields
that the determinant of the matrix ${\cal D}$ is equal to \begin{equation}
\mbox{Det}
\left [{\cal D} \right ]=\left ( p_0^2 - p_1^2 - p_2^2 - p_3^2 -m^2 \right
)^2 \frac{( p_0^2 -p_1^2 -p_2^2 -p_3^2 +3m^2 +4mp_0 )^2}{16m^4 (p_0 +m)^4}
\quad.
\end{equation}
As a result of equating the determinant to zero we
deduce that the equation (\ref{genweq1}) has eight solutions in total with
\begin{equation}
p_0 = \pm \sqrt{{\bf p}^{\,2} +m^2}\quad,
\end{equation}
each two times; and with the
acausal dispersion relations:
\begin{equation}
p_0 = -2m \mp \sqrt{{\bf p}^{\,2}
+m^2}\quad,\label{adr2}
\end{equation}
each two times.

The same situation could be met in deriving the Dirac equation by
the Ryder-Burgard-Ahluwalia technique provided
that we do {\it not} apply to the constraint $p_0^{\,2} -{\bf p}^{\,2}
=m^2$ from the beginning. Indeed,
\begin{mathletters} \begin{eqnarray}
\Lambda_{_R} (p^\mu \leftarrow \overcirc{p}^\mu)
\Lambda_{_L}^{-1} (p^\mu \leftarrow \overcirc{p}^\mu) &=&
\frac{p_0^2 + 2mp_0 + {\bf p}^2 +m^2 + 2(p_0+m) (\bbox{\sigma}\cdot
{\bf p})}{2m(p_0+m)}\quad,\\
\Lambda_{_L} (p^\mu \leftarrow \overcirc{p}^\mu)
\Lambda_{_R}^{-1} (p^\mu \leftarrow \overcirc{p}^\mu) &=&
\frac{p_0^2 + 2mp_0 + {\bf p}^2 +m^2 - 2(p_0+m) (\bbox{\sigma}\cdot
{\bf p})}{2m(p_0+m)}\quad.
\end{eqnarray}
\end{mathletters}
Thus, the second-order momentum-representation ``Dirac" equation can be
written:
\begin{equation}
{1\over 2m(p_0+m)} \left [ (\gamma^\mu p_\mu
\mp m)\gamma^0 +2m \right ] (\gamma^\nu p_\nu \mp m) \Psi_{\pm} (p^\mu) =0
\quad,\label{deq1}
\end{equation}
or
\begin{equation}
{1\over 2m(p_0+m)} (\gamma^\nu p_\nu \mp m) \left [ \gamma^0
(\gamma^\mu p_\mu
\mp m) +2m \right ] \Psi_{\pm} (p^\mu) =0\quad.\label{deq2}
\end{equation}
The corresponding coordinate-representation of these equations
($m\neq 0$ and $p_0 \neq -m$) is
\begin{equation}
\left [ (i\gamma^\mu \partial_\mu
-m)\gamma^0 + 2\wp_{u,v} m \right ] (i\gamma^\nu \partial_\nu - m) \Psi
(x^\mu) =0\quad,
\end{equation}
or
\begin{equation}
(i\gamma^\mu \partial_\mu -m)
\left [ \gamma^0 (i\gamma^\mu \partial_\mu
-m) +2\wp_{u,v} m \right ] \Psi (x^\mu) =0\quad,
\end{equation}
where $\wp_{u,v} =\pm 1$ depending on what solutions, with either positive
or negative energies, are considered.

What about the equation
(\ref{genweq1})?  Can it be put in a more convenient form? The
eight-component form, we proposed recently~[2d,Eqs.(17,18)],
does not have acausal solutions. In the process of its deriving we
have assumed
certain relations\footnote{See, {\it e.g.}, formulas (48) of
ref.~\cite{DVA96}.} between $\lambda^{S,A} (p^\mu)$ and $\rho^{S,A}
(p^\mu)$.  In the present article we are not going to apply them.
Following the procedure of deriving the equations (\ref{deq1},\ref{deq2})
one can arrive at the rather complicated equation:
\begin{eqnarray}
\lefteqn{{1\over 4m (p_0 +m)}
\left \{ (\gamma^\mu p_\mu + m\gamma^0)\left [ {\cal S}
(\gamma^\nu p_\nu +m\gamma^0)
-2m\gamma^0\right ] + \right.}\nonumber\\ &+&\left.  \left [
(\gamma^\mu p_\mu
+m\gamma^0) {\cal S} -2m\gamma^0 \right ] (
\gamma^\nu p_\nu +m\gamma^0 )\right \}
\lambda^{S,A} (p^\mu) =0 \quad,\label{td}
\end{eqnarray}
where
\begin{equation} {\cal S} =
\pmatrix{0 & \zeta_\lambda \Theta \Xi\cr \zeta_\lambda \Xi^{-1} \Theta &
0\cr}\quad.
\end{equation} But, as mentioned in~\cite{DVA96}, one may
consider that $\phi$ in the generalized Ryder-Burgard relation (see Eq.
(27) of ref.~\cite{DVA96} or Eq. (38) in~[2c]) is the azimuthal angle {\it
associated with} ${\bf p}$, the 3-momentum of the particle. In this case
one can find commutation relations between $\hat p \equiv \gamma^\mu
p_\mu$, matrices $\gamma^5$, $\gamma^0$ and ${\cal S}$.
\begin{equation}
\left [ \hat p , {\cal S} \right
]_{-} =0\quad,\quad \left [\gamma^0 , {\cal S}\right ]_- =0\quad,\quad
\left [ \gamma^5 , {\cal S} \right ]_+ = 0\quad,    \label{cr}
\end{equation}
and
\begin{equation} {\cal S}
\lambda^{S,A} (p^\mu) = \lambda^{S,A} (p^\mu) \quad,\label{sa}
\end{equation}
because
in this case
\begin{equation}
\Lambda_{_{L,R}}^\ast = \mit{\Xi}
\Lambda_{_{L,R}} \mit{\Xi}^{-1}\quad.
\end{equation}
We finally arrive at
\begin{equation} \left [\hat p^2 -m^2 \right ] \openone_{4\times
4}\,\,\lambda^{S,A} (p^\mu) = 0\quad,
\end{equation}
{\it i.e.}, at the
Klein-Gordon equation for each component of $\lambda^{S,A} (p^\mu)$. Why
did acausal solutions fall out?  It appears bispinors $\lambda^A
(p^\mu) \equiv -\gamma^5 \lambda^{S} (p^\mu)$ can satisfy the
positive-energy equation ($\zeta_\lambda =i$) and bispinors
$\lambda^S \equiv -\gamma^5 \lambda^A (p^\mu)$, the negative-energy one
($\zeta_\lambda =-i$), but dispersion relations
will be acausal, Eq. (\ref{adr2}), in this non-ordinary case.\footnote{In
the process of the proof one should take into account commutation
relations (\ref{cr}) and hence that ${\cal S}^{^+}
\gamma^5 \lambda^S (p^\mu) = \lambda^A (p^\mu)$ and ${\cal S}^{^-}
\gamma^5 \lambda^A (p^\mu) = \lambda^S (p^\mu)$.} So, assuming that
in the equations (\ref{td})  one should take $\zeta_\lambda
=i$ for describing $\lambda^S$ and $\zeta_\lambda = -i$, for $\lambda^A$
we, in fact, implicitly impose mass-shell constraints.
The same situation is for the equations (\ref{deq1},\ref{deq2}),
$u (p^\mu)$ and $v (p^\mu)\equiv \gamma^5 u (p^\mu)$ can satisfy both the
positive- and the negative-energy equations, but the dispersion relations
could be unusual.

From a mathematical viewpoint the origin of  appearance of these
solutions seems to be related with the properties with respect to
herimitian conjugation operation of the Lorentz
transformation operators,  see~\cite[p.404]{DVAT} for discussion. One
should further note that the problem of acausal solutions  have
intersections with a mathematical possible situation when operators of the
continuous Lorentz transformations are combined  with other
transformations of the Poincar\`e group to give $\Lambda_{_R} = -
\Lambda_{_L}^{-1}$.  Thus, the question, whether these solutions would
have some physical significance, should be solved on the basis of the
rigorous analysis of the general structure of the Poincar\`e
transformation group and of the experimental situation in neutrino
physics.

Finally, let us mention that another second-order equation in the
$(1/2,0)\oplus (0,1/2)$ representation space has been investigated
in~\cite{Barut}  and relations with the problem of the lepton mass
spectrum have been revealed (see also~\cite{Markov,Nigam}).

\medskip

{\it Acknowledgments.}
I acknowledge the help of Profs. D. V. Ahluwalia, A. E. Chubykalo,
M. W. Evans, A.~F. Pashkov and S. Roy. I am grateful to Zacatecas
University, M\'exico, for a Full Professor position.

This work has been partially supported by el Mexican Sistema
Nacional de Investigadores, el Programa de Apoyo a la Carrera Docente
and by the CONACyT, M\'exico under the research project 0270P-E.


\bigskip
\bigskip
\bigskip

\begin{references}

\footnotesize{
\baselineskip13pt

\bibitem{DVA96} D. V. Ahluwalia, Int. J. Mod. Phys. {\bf 11} (1996) 1855

\bibitem{DVO95a} V. V. Dvoeglazov, Rev. Mex. Fis. Suppl. {\bf 41}
(1995) 159;  Bol.  Soc.  Mex.  Fis.  Suppl.  {\bf 9} (1995) 28;
Int.  J.  Theor.  Phys.  {\bf 34} (1995) 2467; Nuovo Cim. {\bf
108}A (1995) 1467

\bibitem{Ziino} G. Ziino, Ann. Fond. L. de Broglie {\bf 14} (1989) 427;
ibid {\bf 16} (1991) 343; Int. J. Mod. Phys. {\bf 11} (1996) 2081;
A.  O.  Barut and G.  Ziino, Mod.  Phys.  Lett.
A{\bf 8} (1993) 1011

\bibitem{DVO95b} V. V. Dvoeglazov, Nuovo Cim. {\bf 111}B (1996) 483

\bibitem{NO} C. Athanassopoulos {\it et al.}, Phys. Rev. Lett. {\bf 75}
(1995) 2650; S. M. Bilen'ky {\it et al.}, Phys. Lett. B{\bf 356} (1995) 273

\bibitem{Maj} E. Majorana,  Nuovo Cim.  {\bf 14} (1937) 171

\bibitem{Markov} M. A. Markov, ZhETF {\bf 7} (1937) 579, 603;
Preprint JINR D-1345, Dubna, 1963

\bibitem{MLC} J. A. McLennan,  Phys. Rev. {\bf 106} (1957) 821;
K. M. Case,  Phys. Rev. {\bf 107} (1957) 307

\bibitem{Fush} N. D. Sen Gupta, Nucl. Phys.
B{\bf 4} (1967) 147;  Z. Tokuoka, Prog. Theor. Phys. {\bf 37} (1967)
603; V.  I.  Fushchich and A.  L.  Grishchenko, Lett.  Nuovo Cim.  {\bf 4}
(1970) 927; V.  I.  Fushchich, Nucl. Phys.  B{\bf 21} (1970) 321; Theor.
Math.  Phys.  {\bf 7} (1971) 3; Lett.  Nuovo Cim.  {\bf 4} (1972) 344; M.
T.  Simon, ibid {\bf 2} (1971) 616; T.  S.  Santhanam and A. R.  Tekumala,
ibid {\bf 3} (1972) 190; M.  Seetharaman, M.  T. Simon and P. M.  Mathews,
Nuovo Cim.  {\bf 12}A (1972) 788

\bibitem{DVO95c} V. V. Dvoeglazov, Hadronic J. Suppl. {\bf 10} (1995) 349

\bibitem{RB} L. H. Ryder, {\it Quantum Field Theory.} (Cambridge
University Press, 1987)

\bibitem{BWW} D. V. Ahluwalia, M. B. Johnson and T. Goldman, Phys. Lett.
B{\bf 316} (1993) 102

\bibitem{DVO95d} V. V. Dvoeglazov, {\it De Dirac a Maxwell: Un Camino Con
Grupo de Lorentz.} Investigaci\'on Cient\'{\i}fica, in press

\bibitem{DVAT} D. V. Ahluwalia and D. J. Ernst, Int. J. Mod. Phys. E{\bf
2} (1993) 397

\bibitem{Barut} A. O. Barut, Phys. Lett. {\bf 73}B (1978) 310

\bibitem{Nigam} R. Acharya and B. P. Nigam, Lett. Nuovo Cim. {\bf 31}
(1981) 437; B. P. Nigam, J. Phys. G{\bf 16} (1990) 1553

}

\end{references}
\end{document}